\documentstyle[epsbox,preprint]{jpsj} 
\title
{
Variational Monte Carlo Study on the Superconductivity 
in the Two-Dimensional Hubbard Model
}
\author
{
Kunihiko {\sc Yamaji}$^{a,b}$\footnote{
Corresponding author, Electrotechnical Laboratory, 
1-1-4 Umezono, Tsukuba 305-8568, Japan; tel: 81-298-54-5368, fax: 81-298-54-5099,
e-mail: yamaji@etl.go.jp}, 
Takashi {\sc Yanagisawa}$^{a}$, 
Takeshi {\sc Nakanishi}$^{a}$\footnote{
Present address: Institute of Physical and Chemical Research,
2-1 Hirosawa, Wako 351-0198, Japan}
and Soh {\sc Koike}$^{b,a}$
}
\inst
{
$^a$ Electrotechnical Laboratory,
1-1-4 Umezono, Tsukuba, Ibaraki 305-8568, Japan \\
$^b$ Institute of Materials Science, University of Tsukuba,
Ten-nodai, Tsukuba 305-8573, Japan
}
\recdate
{May 29, 1998} 
\abst
{
The possibility of superconductivity (SC) in the ground state of the
two-dimensional (2D) Hubbard model was investigated 
by means of the
variational Monte Carlo method. 
The energy gain of the $d$-wave SC state, 
obtained as the difference of the
minimum energy with a finite gap and that with zero gap, was
examined with respect to dependences on $U$, electron density 
$\rho$ and next nearest neighbor transfer $t'$
mainly on the $10 \times 10$ lattice.
It was found to be maximized around $U = 8$ 
(the energy unit is nearest neighbor transfer $t$).
It was shown to sharply increase for negative values of $t'$
and have a broad peak for
$t' \sim -0.10$. For these value of $t'$ the
energy gain was a smooth increasing function of $\rho$ 
almost independent of the shell structure
in the region starting from $\sim 0.76$ up to the upper bound
of investigation 0.92.
This clearly indicates that
the result is already close to the value in the bulk limit.
For $t' = 0$, the energy gain
depended on the electronic shell state. 
This suggests the $10 \times 10$ lattice
is not sufficiently large for this case,
although it is highly plausible that 
the bulk limit value is finite.
Competition between the SC and the
commensurate SDW states was also investigated. 
When $t' = 0$, the ground state is SDW 
in the range of $\rho \geq \sim 0.84$. 
The SC region slightly extends up to $\sim 0.87$ 
for $t' \sim -0.10$.
Consequently the present results strongly support 
an assertion that the 2D Hubbard model with $t' \sim -0.1$
drives SC by itself in the $\rho$ region from $\sim 0.76$
to $\sim 0.87$. 
The above features are in a fair agreement with the 
phase diagram of the optimally and overly hole-doped cuprates.
The energy gain in the SC state with suitable parameters
is found to be in reasonable agreement with the 
condensation energy in the SC state of YBa$_2$Cu$_3$O$_7$.
The corresponding $t$-$J$ model proves to give 
an order-of-magnitude larger energy gain,
which questions its validity. \\
 \\
\verb+PACS codes: 74.20.Mn, 74.25.Bt, 74.25.Dw, 74.72.-h+
}

\kword
{
two-dimensional Hubbard model, 
variational Monte Carlo method, 
$d$-wave superconductivity, SDW,
next nearest neighbor transfer, condensation energy
}

\begin{document}
\sloppy
\maketitle

\section{Introduction}

Recently the mechanisms of superconductivity (SC) in 
high-temperature cuprate superconductors and organic superconductors
have been extensively studied using various two-dimensional (2D)
models of electronic interactions. The 2D Hubbard model is the
simplest and the most fundamental one among such models. 
Early studies of this model 
using the quantum Monte Carlo (M.C.) method
indicated the existence of an
attractive interaction for SC [\citen{rf:hirsch}].
However, this possibility has been controversial. 
Some authors asserted from quantum M.C.
results that the enhanced SC correlation does not develop into the
predominant one at low temperatures or in the ground state of this
model [\citen{rf:white,rf:imada,rf:scalapino,rf:dagotto}]. 
Some authors supported this possibility by numerical 
results using the variational Monte Carlo (M.C.) method [\citen{rf:gia}].
Since it is of prime importance to establish the simplest
electronic model for superconductivity 
and since these early studies had shortcomings
due to the restrictions to the investigated parameter space,
the problem deserves more extensive
investigation from various aspects such as dependences 
on the system size, the on-site Coulomb energy $U$, 
the electron density $\rho$ and
the next nearest neighbor transfer energy $t'$.
Recently, appropriate values of $t'$ were found 
to remarkably enhance SC correlations [\citen{rf:f2,rf:f3,rf:huss}].
We apply the variational Monte Carlo method [\citen{rf:yoko-shiba,rf:gros}] mainly to the 
$10 \times 10$-site system with electron numbers from 68 to 92.
This method has a merit that it allows to treat larger
values of $U$ than the quantum M.C. methods. 
We revisit the issue investigated in ref. [\citen{rf:gia}],
carrying out more extensive study
with higher precision in a wider parameter space.
In a previous preliminary report [\citen{rf:naka}] 
we worked out a variational Monte Carlo calculation 
84 electrons on the $10 \times 10$ lattice and 
confirmed that 
the 2D Hubbard model has the $d$-wave SC state.
In the present work first we examine the $U$-dependence of 
the energy gain due to the condensation into the SC state 
and find the optimal
value of $U$ is about $8$ in units of the nearest neighbor 
transfer energy $t$.
In the previous report we confirmed
that the energy gain in the SC state sharply increases with the
introduction of $t'$ with a negative value. 
In this report we calculate the energy gain as a function of $t'$ in a sufficiently wide range
for a fixed value of $U = 8$ and show 
that it has a broad peak
around a negative value $t' \sim -0.1$. 
The approximate peak energy gain in the SC state, 
i.e., at $t' \sim - 0.1$, is plotted as a function of electron
density $\rho$ together with that for $t' = 0$.
When $t' \sim -0.1$, the energy gain starts to be finite 
at about $\rho = 0.68$ and increases fairly 
smoothly with increase
of $\rho$ in the range of $0.76 \leq \rho \leq 0.92$.
We carried out calculations in both cases of 
open and closed shells.
The smoothness of the energy gain against $\rho$ indicates that
the dependence of the energy gain on the shell structure is weak,
in contrast to the case with $t' = 0$, and 
that the results are already close
to the bulk limit. This means that the 2D Hubbard model with
$t' \sim -0.1$ has a definite bulk-limit 
energy gain per site in the SC state.
While when $t' = 0$, the energy gain 
in the closed shell state was found
much smaller than that in the open shell state. This indicates that
the $10 \times 10$ lattice is not sufficiently large 
for the case of $t' = 0$, contrary to the case of $t' \sim -0.1$, although a kind of average of the results in both shell states 
is very probable to give a finite 
energy gain per site in the bulk limit.
Of course it is needed to treat larger lattices with $t' = 0$ 
but some important findings
at the present stage are considered to deserve a report,
with the latter problem left to be settled,
since with larger sizes computation needs extremely long time. 

Thirdly, we calculate the energy gain
in another ordered state, i.e., spin density wave (SDW) state. 
We treat the SDW with a fixed wave vector $(\pi, \pi)$. 
Here the length unit is the lattice constant.
The energy gain in the SDW state was found to quickly drop 
with decrease of $\rho$ from unity and to vanish at $\rho$ just 
below 0.84 when $t' = 0$.
However, at $\rho = 0.84$ the SDW state was slightly 
more stable than the SC state 
so that the boundary between both states lies just below 
$\rho = 0.84$ for $t' = 0$.
$t'$ is known to destabilize the SDW in the hole-doped state, 
while it promotes SC pairing. 
We study to what extent the phase boundary between 
the competing SC and SDW states is affected by the 
negative $t'$. 
The boundary is appreciably pushed up to the higher $\rho$
side with increase of the absolute value of the 
negative $t'$, although not largely.  

 We point out that the resultant SC region 
from $\rho \sim 0.76$ to $\rho \sim 0.86$ and the
increased SC energy gain with increase of $\rho$ are 
in accord with 
experimental features of cuprate high-$T_c$
superconductors in the overdoped region.
The observed energy gain in the SC state is found to be in fair
agreement with the experimental condensation energy estimated from the
critical magnetic field $H_c$ and the specific heat of YBa$_2$Cu$_3$O$_7$.
On the other hand the $t$-$J$ model as an effective Hamiltonian of 
the Hubbard model is noticed to give 
an enormous overestimation of
the condensation energy.

 In the next section the model and the method are described.
Results on the SC and SDW ground states are given in \S3 and \S4,
respectively. In \S5 the obtained results are compared with experiments
and also with other computational works. The final section gives the 
summary.
Short reports on the present results were given in [\citen{rf:naka,rf:yama}].

\section{Model and Method}

     Our model is the 2D  Hubbard model defined by 
\begin{equation}
H =
-t \sum_{<jl>,\sigma} (c_{j\sigma}^{\dag}c_{l\sigma} + 
                               \mbox{H.c.})
+ U \sum_j c_{j\uparrow}^{\dag} c_{j\uparrow}
                                  c_{j\downarrow}^{\dag} c_{j\downarrow},
                             \label{eq:H}
\end{equation}
where $c_{j\sigma}^{\dag}$ ($c_{j\sigma}$) is the creation (annihilation) 
operator of an electron with spin $\sigma$ at the $j$th site;
the sites form a rectangular lattice;
$t$ is the transfer energy between the nearest-neighbor
(n.n.) sites; $t$ is our energy unit;
$<jl>$ denotes summation over all the n.n. bonds.
$U$ is the on-site Coulomb energy.  
In this report, we also study the effect of 
$t'$ between n.n.n. sites by including
\begin{equation}
H_{nnn} = -t' \sum_{<<jl>>,\sigma} (c_{j\sigma}^{\dag}c_{l\sigma} + 
                               \mbox{H.c.}) \label{eq:H2} 
\end{equation}
in the Hamiltonian; in the above equation
$<<jl>>$ means summation over the n.n.n. pairs.

     Our trial wave function is a
Gutzwiller-projected BCS-type wave function defined as [\citen{rf:yoko-shiba,rf:gros}]:
\begin{eqnarray}
\Psi_{\mbox{\scriptsize s}} &=& P_{N_e} P_{\mbox{\scriptsize G}}
                                      \psi_{\mbox{\scriptsize BCS}}, \\
\psi_{\mbox{\scriptsize BCS}} &=& \prod_{{\mib k}} (u_{\mib k} + v_{\mib k}
        c_{{\mib k}\uparrow}^{\dag}c_{-{\mib k}\downarrow}^{\dag})|0>,
\end{eqnarray}
where $P_{\mbox{\scriptsize G}}$ is the Gutzwiller projection operator 
given by
\begin{equation}
P_{\mbox{\scriptsize G}} = \prod_j [1 - (1-g)n_{j\uparrow}n_{j\downarrow}]; 
\end{equation}
$g$ is a variational parameter in the range from 0 to unity
and $j$ labels a site in the real space. 
$P_{N_e}$ is a projection operator
which extracts only the states with a fixed total electron number 
$N_e$.  Coefficients $u_{\mib k}$ and $v_{\mib k}$
appear in our calculation only in the ratio defined by
\begin{eqnarray}
& & v_{\mib k}/u_{\mib k} =  \Delta_{\mib k}/( \xi_{\mib k}
       + \sqrt{ \xi_{\mib k}^2 + \Delta_{\mib k}^2 } ), \label{eq:vu} \\
\xi_{\mib k} &=& - 2t (\cos k_x + \cos k_y) 
- 4t' \cos k_x \cos k_y - \mu , \label{eq:ek}
\end{eqnarray}
where 
$\Delta_{\mib k}$ is a      
${\mib k}$-dependent gap function defined later;
$\mu$ is a variational parameter working like the chemical
potential in the trial wave function;
$c_{{\mib k}\sigma}$ is the Fourier transform of $c_{j\sigma}$.
Neglecting constant factors, $\Psi_{\mbox{\scriptsize s}}$ can be
rewritten as
\begin{eqnarray}
\Psi_s 
&\sim& P_{N_e} P_{\mbox{\scriptsize G}}
        \exp [\sum_{\mib k} (v_{\mib k} / u_{\mib k})
  c_{{\mib k}\uparrow}^{\dag}c_{-{\mib k}\downarrow}^{\dag}]|0>, \label{eq:PP} \\ 
&=& P_{N_e} P_{\mbox{\scriptsize G}}
        \exp [\sum_{j,l} a(j,l) 
      c_{j\uparrow}^{\dag}c_{l\downarrow}^{\dag}]|0>, \\ 
& \sim\ & P_{\mbox{\scriptsize G}} [\sum_{j,l} a(j,l) 
      c_{j\uparrow}^{\dag}c_{l\downarrow}^{\dag}]^{N_e/2}|0>, \\ 
&=& P_{\mbox{\scriptsize G}}
      \sum_{j_1,..,j_{N_e/2},l_1,..,l_{N_e/2}}
      A(j_1, .., j_{N_e/2};l_1, .., l_{N_e/2})  \nonumber  \\
&\times& c_{j_1 \uparrow}^{\dag} c_{j_2 \uparrow}^{\dag} ...
c_{j_{N_e/2} \uparrow}^{\dag} c_{l_1 \downarrow}^{\dag} 
c_{l_2 \downarrow}^{\dag} ... c_{l_{N_e/2} \downarrow}^{\dag} |0>,
\end{eqnarray}
where $a(j,l)$ is defined by
\begin{equation}
a(j,l) = (1/N_s) \sum_{{\mib k}} (v_{\mib k} / u_{\mib k})
                     \exp [\mbox{i} {\mib k} \cdot ({\mib R}_l - 
                      {\mib R}_j )], \label{eq:ajl}
\end{equation}
with $N_s$ being the number of sites and
\begin{eqnarray}
& & A(j_1, j_2,..., j_{N_e/2};l_1, l_2,..., l_{N_e/2}) = \nonumber  \\
& &
\left|
   \begin{array}{cccc}
       a(j_1,l_1) & a(j_1,l_2)&\ldots & a(j_1,l_{N_e/2}) \\
       a(j_2,l_1) & a(j_2,l_2)&\ldots & a(j_2,l_{N_e/2}) \\
       \multicolumn{4}{c}{\dotfill}\\
       a(j_{N_e/2},l_1) & a(j_{N_e/2},l_2)&\ldots & a(j_{N_e/2},l_{N_e/2}) \label{eq:det} \\
   \end{array}
\right| ;\ 
\end{eqnarray}
$j_1,j_2, \ldots$ are arranged in the ascending order; so are
$l_1, l_2, \ldots$.
Then the ground state energy 
\begin{equation}
E_g = < H > \equiv <\Psi_s| H |\Psi_s> / <\Psi_s|\Psi_s> \label{eq:average}
\end{equation}
is obtained using a M.C. procedure [\citen{rf:yoko-shiba,rf:gros}].
In order to minimize computation time in the M.C. computation 
the values of the cofactors of the matrix in 
eq. (\ref{eq:det}) were stored and corrected
at each time when the electron distribution was modified. 
Errors accumulated in the cofactors
after corrected many times were avoided by  
recalculating cofactors as determinants, instead of getting 
them by correction, 
after a certain number of M.C. steps.
We tested our programs
using exact diagonalization results for small systems.
We optimized $E_g$ with respect to 
$g$, $\Delta_{\mib k} $ and $\mu$.
In the later stage of work 
we employed the correlated measurements method [\citen{rf:wil,rf:koba}]
in the process of searching optimal parameter
values minimizing $E_g$.
It enables to precisely locate the minimum position.

\section{Results for the Superconducting Ground State}

\subsection{Case of Simple 2D Hubbard Model}

     We studied the cases of the $d$-,
extended $s$- ($s^*$-) and $s$-wave gap functions as follows:
\begin{eqnarray}
\begin{array}{ccccc}
d & \Delta_{\mib k} &=& \Delta (\cos k_x - \cos k_y ),
\label{eq:dw}
      \\
s^* & \Delta_{\mib k} &=& \Delta (\cos k_x + \cos k_y ),
     \\
s & \Delta_{\mib k} &=& \Delta
\end{array}
\end{eqnarray}
The sizes of the lattice we treated are $6 \times 6$ and 
$10 \times 10$
having electron density close to unity with slight hole
doping into the half-filled state.

        Results indicating the occurrence of the $d$-wave
superconductivity were obtained even for the case of
$N_e = 32$ on the $6 \times 6$ lattice with the
periodic and the antiperiodic boundary conditions (b.c.'s)
for the $x$- and the $y$-direction, respectively, 
with $U = 8$ and $t' = 0$.
This set of b.c.'s was chosen so that $\Delta_{\mib k}$ does not
vanish for any ${\mib k}$-points possibly occupied by electrons;
otherwise zero division occurs in eq. (\ref{eq:ajl}).
$E_g/N_s$ is minimized at $\Delta \sim 0.10$. Here $g = 0.3038$
and $\mu = - 0.48$. The energy gain due to the SC gap formation,
i.e., SC condensation energy,  
was estimated at $\sim 0.00028$/site from the difference
between the minimum and the intercept of the 
$E_g$-versus-$\Delta$ curve with
the vertical axis.

\begin{figure}
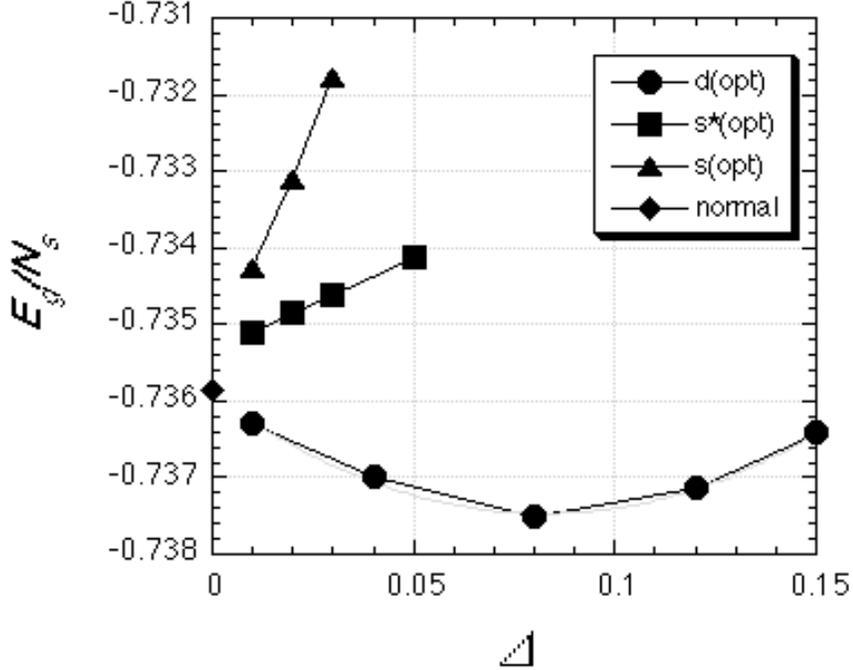

\hspace*{22mm}
\psbox[width=12cm]{fig1.EPSF}
\caption{Computed ground state energy per site $E_g/N_s$ is plotted against $\Delta$ for the case of 84 electrons on the $10 \times 10$ lattice with $U = 8$ and $t' = 0$. Filled circles are for the $d$-wave gap function with $g$ and $\mu$ optimized for each $\Delta$. Filled squares and triangles are for the $s^*$- and $s$-wave gap functions, respectively. The diamond shows the normal state value. Straight lines between data points are the guide for 
the eye. The thin curve is a parabola given by the mean-squares
fit to the $d$-wave data.}
\label{fig:DEg-D}
\end{figure}

\begin{figure}
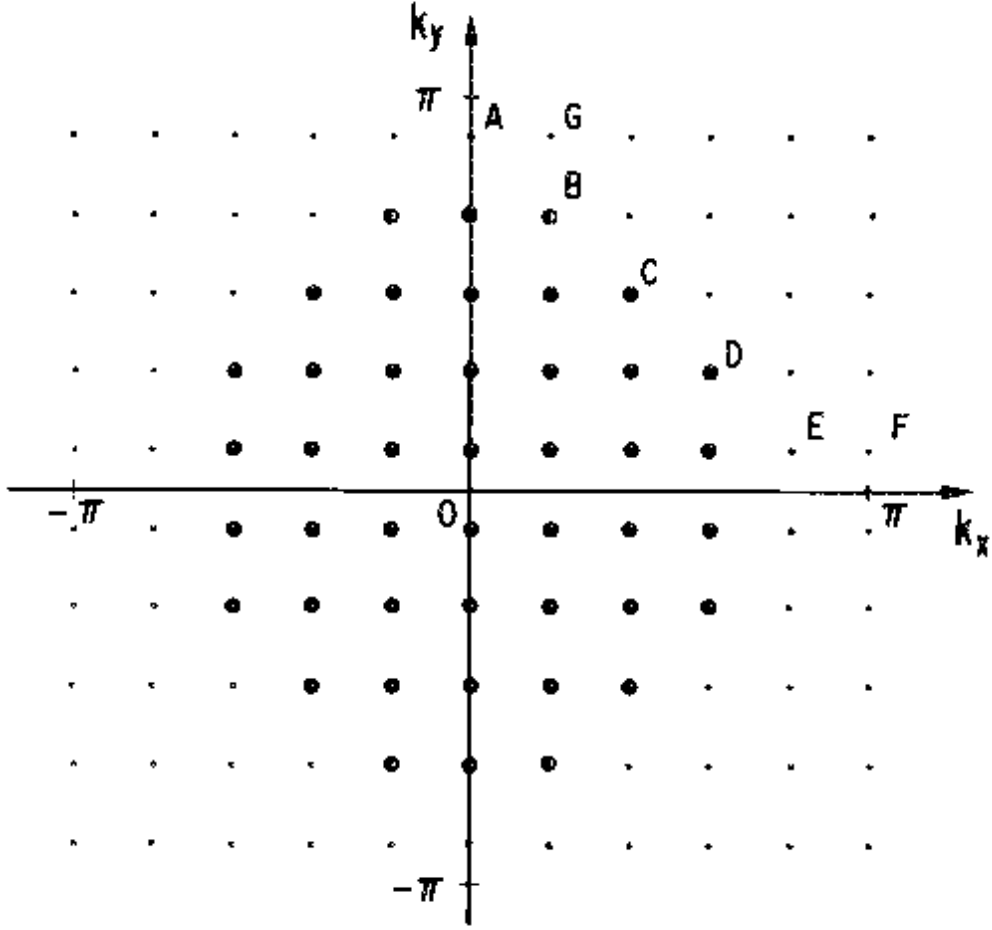

\hspace*{1.2cm}
\psbox[width=14cm]{fig2.EPSF}
\caption{Distribution of ${\mib k}$-points in the reciprocal lattice in the case of the $10 \times 10$ lattice with the boundary conditions mentioned in the text. Closed and empty circles denote doubly occupied and empty sites, respectively, and circles partially closed are partially occupied sites in the case of 84 electrons. A, B, ..., G label ${\mib k}$-points lying in the neighborhood of the Fermi energy for later use.}
\label{fig:kpoints}
\end{figure}

Main results written in this report were obtained 
for the $10 \times 10$ lattice with the same boundary conditions.
Calculated energies per site for
several types of wave functions with $N_e = 84$ on 
the $10 \times 10$ lattice 
are shown in Fig. \ref{fig:DEg-D} for the case of $U = 8$ and $t' = 0$.
Here $E_g/N_s$ is plotted as a function of $\Delta$ for the three
types of gap functions given in eq. (\ref{eq:dw}). 
With the lattice, the b.c.'s and $N_e = 84$,
the electronic shell structure in the limit of 
$U = 0$ is open, i.e., some ${\mib k}$-points are partially filled,
as is illustrated in Fig. \ref{fig:kpoints},
which displays the electron occupancy at the ${\mib k}$-points.
At each value of $\Delta$
in Fig. \ref{fig:DEg-D},
$g = 0.30$ was chosen as the initial value
of $g$ and then the optimal value of $\mu$ was found by the least
squares fit of $E_g$ as a function of $\mu$ to a parabola. 
Using this value of $\mu$, $g$ was optimized again. 
Since $E_g$ was a smooth function of $g$, 
the obtained optimal $g$ and $\mu$ are sufficiently accurate. 
Using these parameter values, $E_g$ was obtained as the
average of the results of eight M.C. calculations each with
$5 \times 10^7$ steps at $\Delta = 0.01,0.04$ and $0.08$
for the $d$-wave.
The standard deviations were less than $0.00011$. At other
points, the numbers of M.C. calculations and steps were
different but their error bars were within $0.00015$ with
the total M.C. steps greater than $2\times 10^8$. 
The diamond shows the normal state value, 
$-0.73585 \pm 0.00024$, obtained from 20 M.C. 
calculations each with $10^7$ steps, 
using the Gutzwiller-projected normal state
wave function.
In the employed normal state $(\pi/6, 7\pi/10)$ and $(-\pi/6,
-7\pi/10)$ are fully occupied but $(\pi/6, -7\pi/10)$ and $(-\pi/6,
7\pi/10)$ are unoccupied
in the component of the trial wave function
prior to the Gutzwiller projection.

       Clearly, $E_g/N_s$ is minimum at a finite value of 
$\Delta \approx 0.08$ in the case of the $d$-wave gap parameter. 
The optimal parameter values at $\Delta = 0.08$ are 
$g = 0.3037$ and $\mu = - 0.4263$. 
The least squares fit of a parabola to $E_g/N_s$ as a function 
of $\Delta$ is of good quality, as seen in Fig. \ref{fig:DEg-D}
and gives the minimum at $\Delta \cong 0.082$. 
The curves of $E_g/N_s$ for the $s$- and $s^*$-wave 
gap functions have definite positive slopes at 
small $\Delta$ and are extrapolated for $\Delta = 0$ to
$\sim -0.7354$ and $\sim -0.7353$, respectively, 
which are practically equal. 
These values are slightly higher than the
value $\sim -0.73605$ given by extrapolating the $d$-wave
fitting parabola to $\Delta = 0$. 
The normal state value of $E_g/N_s = - 0.73585$ 
lies between the two groups of extrapolated values. 
The differences are due to a kind of size effect,
as was explained in the previous report [\citen{rf:naka}].
These differences in the case of $10 \times 10$ lattice are 
much smaller than the depth of the minimum of the $d$-wave curve.

        The energy gain per site in the $d$-wave state 
was obtained from the parabola fitting to the 
energy-versus-$\Delta$ curve 
as its depth of the minimum in reference with the intersection of 
the fitting curve with the vertical axis as $\sim 0.00155$/site. 
It is five times larger than that for $N_e = 32$ and $N_e = 36$. 
This suggests that the size effect is considerable in the
$6 \times 6$ system.
This energy gain per site in the case of $N_s = 100$ and
$N_e = 84$ is of the order of magnitude of the BCS superconducting
condensation energy $\sim$ (state density)$\times \Delta^2 \approx 
0.0010$ since the state density per site per spin around this density
is approximately equal to $1/2\pi$ in units of $t$.

\begin{figure}
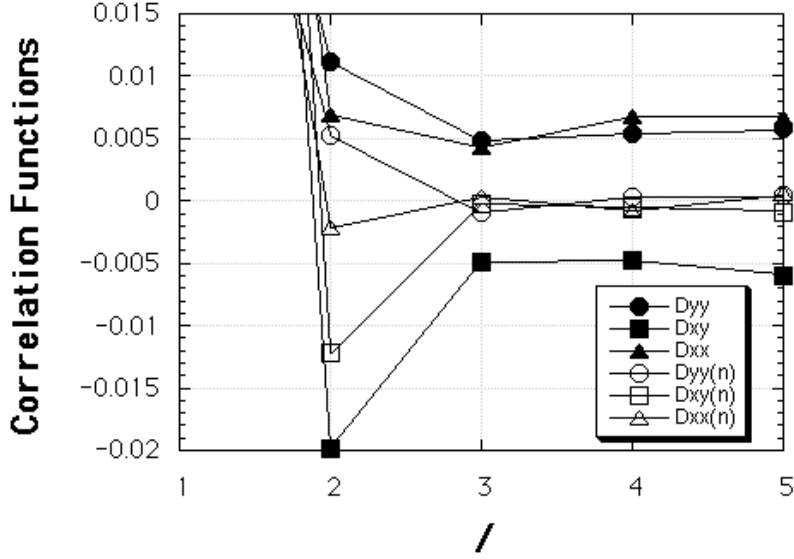

\hspace*{2cm}
\psbox[width=12cm]{fig3.EPSF}
\caption{Parts of the correlation functions $D_{\alpha \beta}$ of superconducting pair operators for case of the $d$-wave minimum state shown in Fig. \ref{fig:DEg-D}. The horizontal axis is the distance $l$ between two positions of two pair operators. Closed symbols are for the superconducting state and open ones for the normal state. $D_{\alpha \beta}$ is defined by eq. (\ref{eq:Dab}).}
\label{fig:cor1}
\end{figure}

        In order to check the superconducting nature of the ground
state with a finite value of $\Delta$, the correlation functions of
BCS pair operators were calculated. Superconducting pair
correlation functions               
$D_{\alpha \beta}(l), \alpha, \beta = x, y$, are defined as:
\begin{equation}
D_{\alpha \beta}(l) = < \Delta_{\alpha}^{\dag}(i+l,j)
           \Delta_{\beta}(i,j) >  , \label{eq:Dab} 
\end{equation}
where $\Delta_{\alpha}(i,j), \alpha = x, y$, denote the annihilation
operators of singlet electron pairs staying on n.n. sites 
as:
\begin{eqnarray}
\Delta_x(i,j) & = & c_{ij\downarrow}c_{i+1,j\uparrow}
                   -  c_{ij\uparrow}c_{i+1,j\downarrow} ,  \\
\Delta_y(i,j) & = & c_{ij\downarrow}c_{i,j+1\uparrow}
                   -  c_{ij\uparrow}c_{i,j+1\downarrow} ,
\end{eqnarray}
where $c_{ij\sigma}$ means the annihilation operator at site $(i, j)$
with spin $\sigma$.
The average $<\ldots>$ is defined in eq. (\ref{eq:average}).
The result for a state close to the minimum $E_g$ point in Fig. 
\ref{fig:DEg-D}, i.e., 
$\Delta = 0.078$, $\mu = -0.428$ and $g = 0.30$, 
is shown in Fig. \ref{fig:cor1}. The vertical scale is enlarged
to highlight the long range parts. 
The correlation extends over the whole lattice as 
expected, showing a clear contrast to the normal state.
The $d$-wave nature appears in the negative sign of $D_{xy}(l)$ for $l
= 2 \sim 5$. 

\begin{figure}
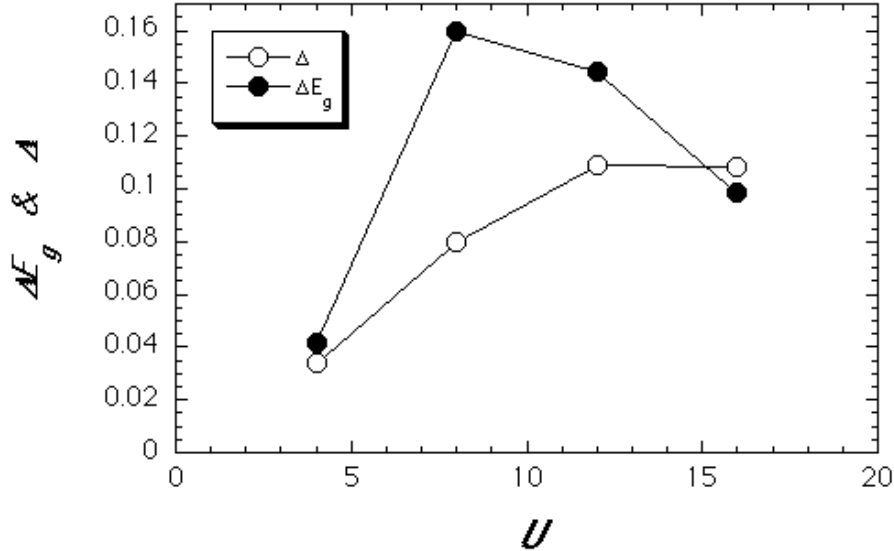

\hspace*{1cm}
\psbox[width=14cm]{fig4.EPSF}
\caption{Energy gain $\Delta E_g$ (closed circles) and gap parameter $\Delta$ (open circles) in the superconducting state are plotted against on-site Coulomb energy $U$ for the same system as in Fig. \ref{fig:DEg-D}.}
\label{fig:DEg-U}
\end{figure}

\subsection{$U$-Dependence}

The $U$-dependence of the energy gain in the SC state was
shown in Fig. \ref{fig:DEg-U}. with the same values 
of $N_s$ and $N_e$. 
The energy gain was determined from the depth of the minimum,
as explained above.
Except for the above-mentioned case of $U = 8$, 
the minimum position defined by optimal $\Delta$, 
$g$ and $\mu$ was determined by means of the correlated measurements 
and $E_g$ was calculated for this position
with the same number of M. C. steps as for $U = 8$.
Next, $E_g$ for $\Delta = 0.01$ was obtained and
then the fitting parabola allowed to get the energy gain. 
The energy gain in this definition gives the condensation energy in 
the SC state in the present approximation.
The energy gain has the
maximum for about $U = 8$. 
It quickly decreases with decrease of $U$. 
It gradually decrease with increase of $U$ over 12. 
In contrast, the value of optimal $\Delta$ is nearly 
proportional to $U$ up to $U$ = 12 and then saturates for
larger $U$.

\subsection{$t'$-Dependence}

     In the 2D Hubbard and the 2D $d$-$p$ models, 
the one-electron state density increases peak-wise around the energy
of the van Hove
singularities located 
in the ${\mib k}$-space region around
$(0, \pi)$ and $(\pi, 0)$. 
When $t'$ is negative 
in the 2D Hubbard model with $H_{nnn}$     
the energy level of van Hove singularity moves toward the Fermi
energy in the hole-doped systems, 
as is seen from eq. (\ref{eq:ek}), which realizes a situation
favorable to get superconductivity for most theories. 
In fact Shimahara and Takada showed that in the RPA framework $T_c$ increases with the introduction of $t'$ [\citen{rf:shima}]. 
Importance of $t'$ in the physics of high-$T_c$ cuprates
was pointed out in [\citen{rf:fuku}].
From the viewpoint of the two-band mechanism of superconductivity
in which the pair-wise transfer of electrons from a certain 
${\mib k}$-region to another ${\mib k}$-region promotes SC, 
the location of the ${\mib k}$-space
region highly contributing to the state density 
in the neighborhood of the Fermi energy 
should lead to a larger energy gain in the SC state [\citen{rf:kondo,rf:y3}]. 
The state density peak is known to be further enhanced with the increase of
electron correlation, as the single-particle dispersion along the lines
from these points to (0, 0) becomes anomalously weak [\citen{rf:shen}].
The increased state density around the singularity has been argued to enhance
$T_c$ of the $d$-wave superconductivity [\citen{rf:naz}].
The value of $t'$ was argued to correlate with the maximum $T_c$ values of cuprate subfamilies [\citen{rf:f2,rf:f3}].
Quantum M. C. studies indicated that $t'$ enables the bulk superconductivity in the 2D Hubbard model [\citen{rf:huss}].

\begin{figure}
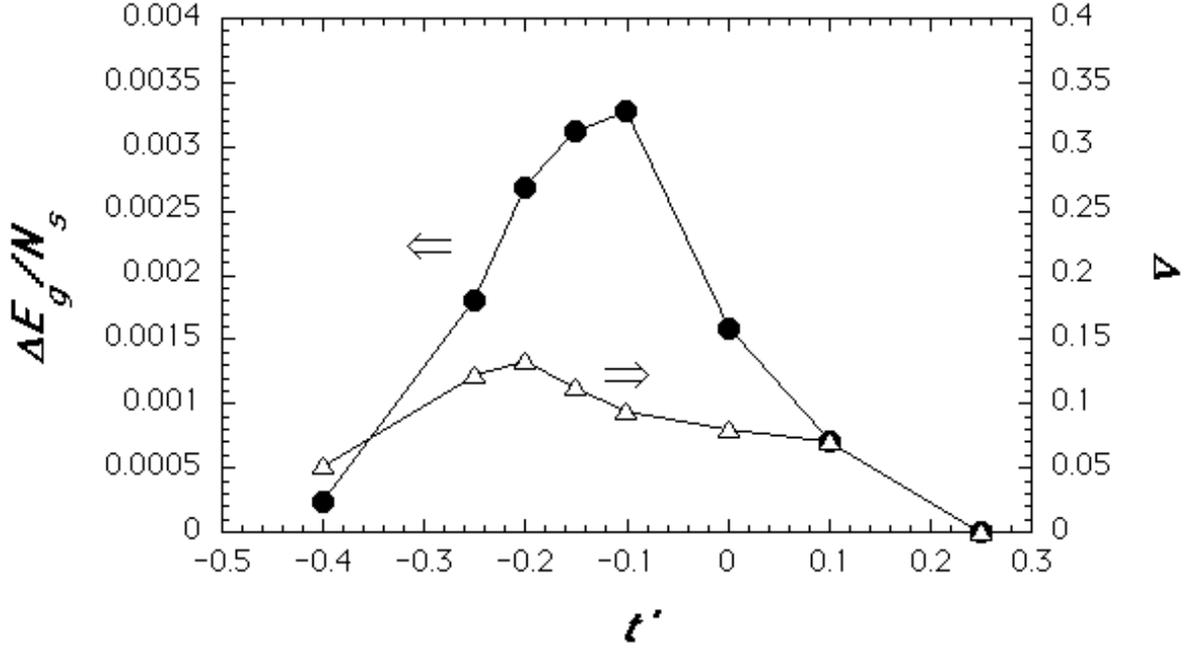

\hspace*{5mm}
\psbox[width=16cm]{fig5.EPSF}
\caption{Energy gain per site $\Delta E_g/N_s$ in the superconducting state is plotted by closed circles with the left vertical scale as a function of next nearest neighbor transfer energy $t'$ in the case of 84 electrons on the $10 \times 10$ lattice with $U = 8$.  Values of $\Delta$ are shown by open triangles with the right-hand-side vertical scale.}
\label{fig:DEg-t'}
\end{figure}

\begin{figure}
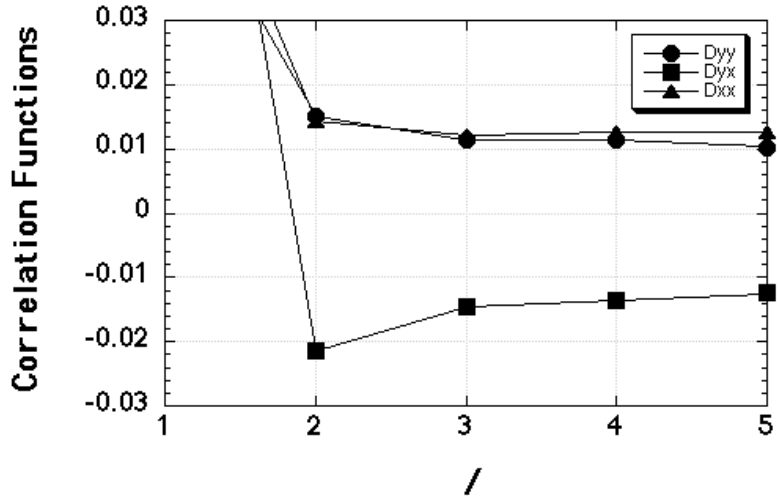

\hspace*{3cm}
\psbox[width=12cm]{fig6.EPSF}
\caption{Parts of the correlation functions of superconducting pair operators for the case of the largest SC energy gain in Fig. \ref{fig:DEg-t'}  with $t' = -0.10$, $\Delta =  0.0941$, $g =  0.3003$, and $\mu =  -0.6152$.  The notation is the same as in Fig. \ref{fig:cor1}. }
\label{fig:cor2}
\end{figure}

\begin{figure}
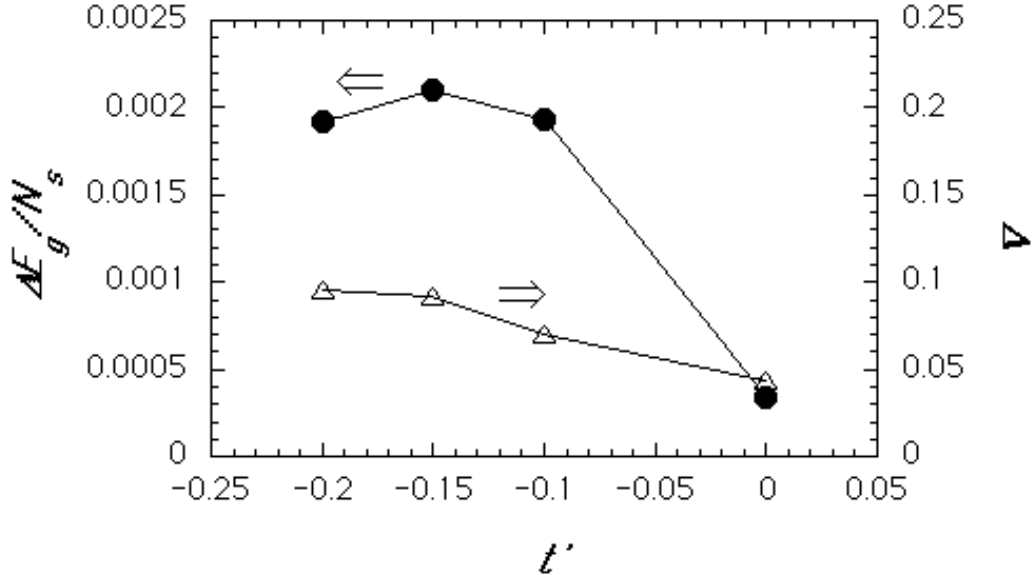

\hspace*{15mm}
\psbox[width=14cm]{fig7.EPSF}
\caption{$\Delta E_g$ and $\Delta$ as functions of $t'$ in the case of $\rho = 0.80$. Other parameter values are the same as in Fig. \ref{fig:DEg-t'}.}
\label{fig:DEg-t'2}
\end{figure}

 We have examined
the $t'$-dependence of the energy gain, 
taking account of $H_{nnn}$ in (\ref{eq:H2})
in our model.  
In the case of $N_e$ = 32 on the $6 \times 6$ lattice, 
the energy minimum became slightly deeper with the change 
of $t'$ from zero to $-0.25$,
but at $t' = 0.25$ the minimum became shallower by a factor of 2. 
In the case of $N_e = 84$ on the $10 \times 10$ lattice 
the energy gain was maximized around 
$t' \cong -0.1 \sim -0.15$ as shown in Fig. \ref{fig:DEg-t'}.
The error bar for $\Delta E_g/N_s$ in the figure is about $0.0003$. 
$\Delta$ is also displayed in the figure with the right-hand-side 
vertical scale.
The error bar for $\Delta$ is about 10 percents of the magnitude.
With a positive value
of $t' = 0.10$, the energy gain quickly decreases, nearly vanishing
when $t' = 0.25$. It also nearly vanishes for $t' \sim -0.40$. 
Figure \ref{fig:cor2} shows the SC
pair correlation functions as functions of the distance between pair
operators in the optimal state for the case of $t' = - 0.10$. 
At the farthest distance their
values are twice larger that those for $t' = 0$ 
(Fig \ref{fig:cor1}), as expected.
Thus, an
appropriate negative value of $t'$ enhances  the $d$-wave
SC in the hole-doped case in the Hubbard model, 
which is in qualitative agreement with the above-mentioned references.
The $t'$-dependences of the energy gain and $\Delta$ in the case of $N_e = 80$ exhibited in Fig. \ref{fig:DEg-t'2} show similar features as in Fig. \ref{fig:DEg-t'}.

\begin{figure}
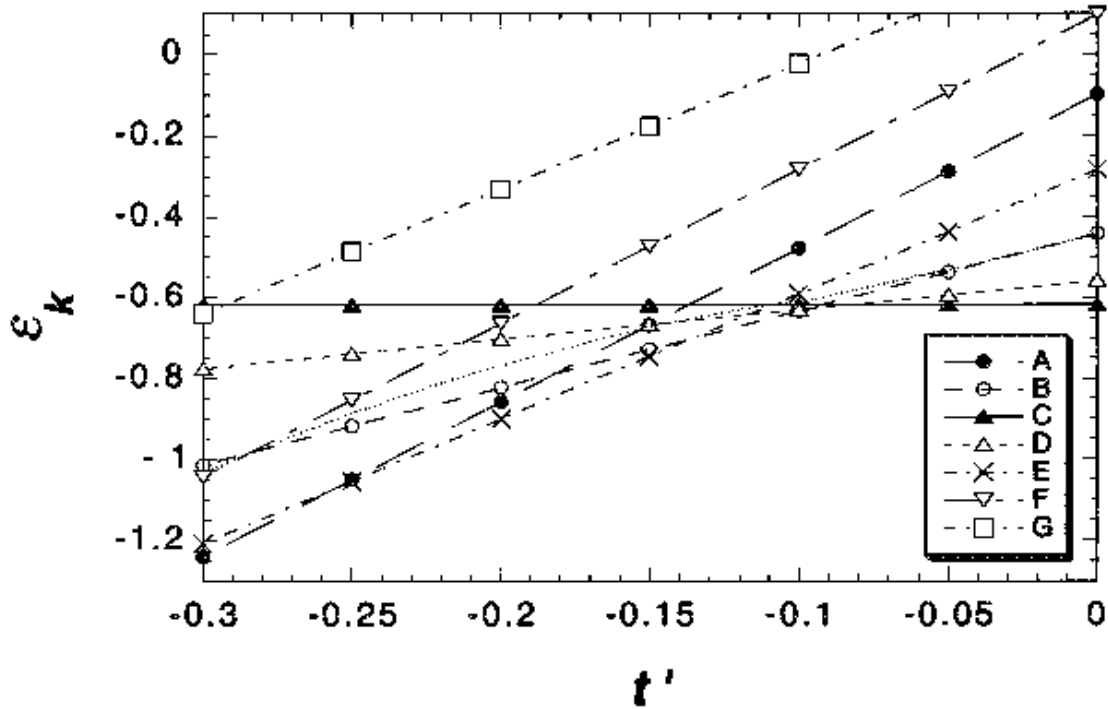

\hspace*{1cm}
\psbox[width=15cm]{fig8.EPSF}
\caption{One-electron levels $\varepsilon_{\mib k}$ of {\mib k}-points A, B, ..., G in the neighborhood of $\varepsilon_{\mbox{\scriptsize F}}$ are plotted as functions of $t'$. Due to orbital and spin degeneracies, each levels can accommodate eight electrons except the A level which can accept four. The dotted curve exhibits $\varepsilon_{\mbox{\scriptsize F}}$ in the case of the $U = 0$ limit.}
\label{fig:levels}
\end{figure}

When one plots one-electron levels (Fig. \ref{fig:levels}),
$\varepsilon_{\mib k} = \xi_{\mib k} + \mu$, for ${\mib k}$-points
in the non-interacting limit as
a function of $t'$, a few occupied levels and a few empty levels are
found to be bunched into a close neighborhood of the 
Fermi energy $\varepsilon_{\mbox{\scriptsize F}}$ 
within $\pm 0.07$ as $t'$ 
takes $-0.1 \sim -0.15$. These levels are located in the 
${\mib k}$-space mainly in the neighborhood of the level of the 
van Hove singularity. Since each
one-electron level is orbitally four- 
(or rarely two-) fold degenerate due to symmetry,
the number of one-electron states whose levels are located in the neighborhood
of $\varepsilon_{\mbox{\scriptsize F}}$ is considerable. 
Since such a feature
is known to enhance the SC pair correlation functions in the 
two-chain Hubbard model [\citen{rf:y3}] 
even in strongly correlated situations, 
this feature is considered to bring about the 
remarkable increase of the energy gain.
Although the particular peaking location
of $t'$ may be slightly size-dependent since the one-electron levels
are size-dependent, such a concentration of one-electron levels
around $\varepsilon_{\mbox{\scriptsize F}}$ is considered to occur 
for any system size when the
appropriate negative $t'$ pushes down the 
van Hove singularity close to $\varepsilon_{\mbox{\scriptsize F}}$.
Therefore, this effect should work even in the bulk limit.

\begin{figure}
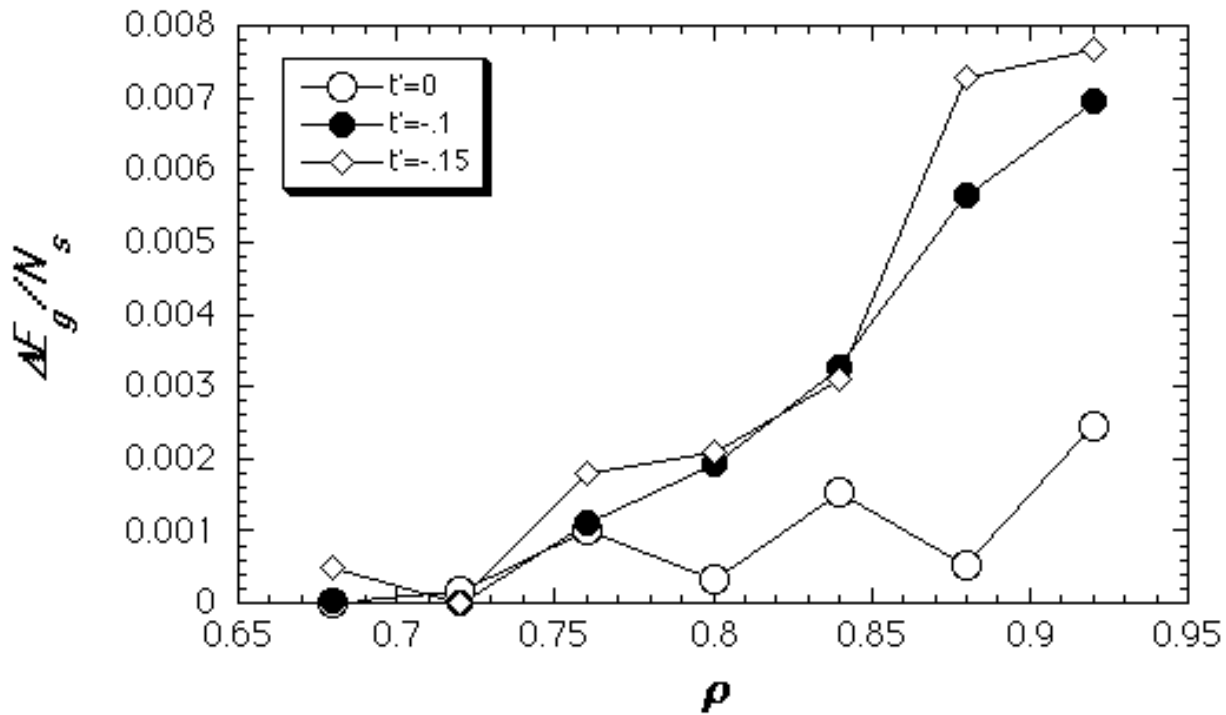

\psbox[width=18cm]{fig9.EPSF}
\caption{Energy gain per site $\Delta E_g/N_s$ in the superconducting state against electron density $\rho$. Open circles are for $t' = 0$, closed ones for $t' = -0.1$ and closed diamonds for $t' = -0.15$. The lattice is $10 \times 10$ and $U=8$. }
\label{fig:gain-r}
\end{figure}

\begin{figure}
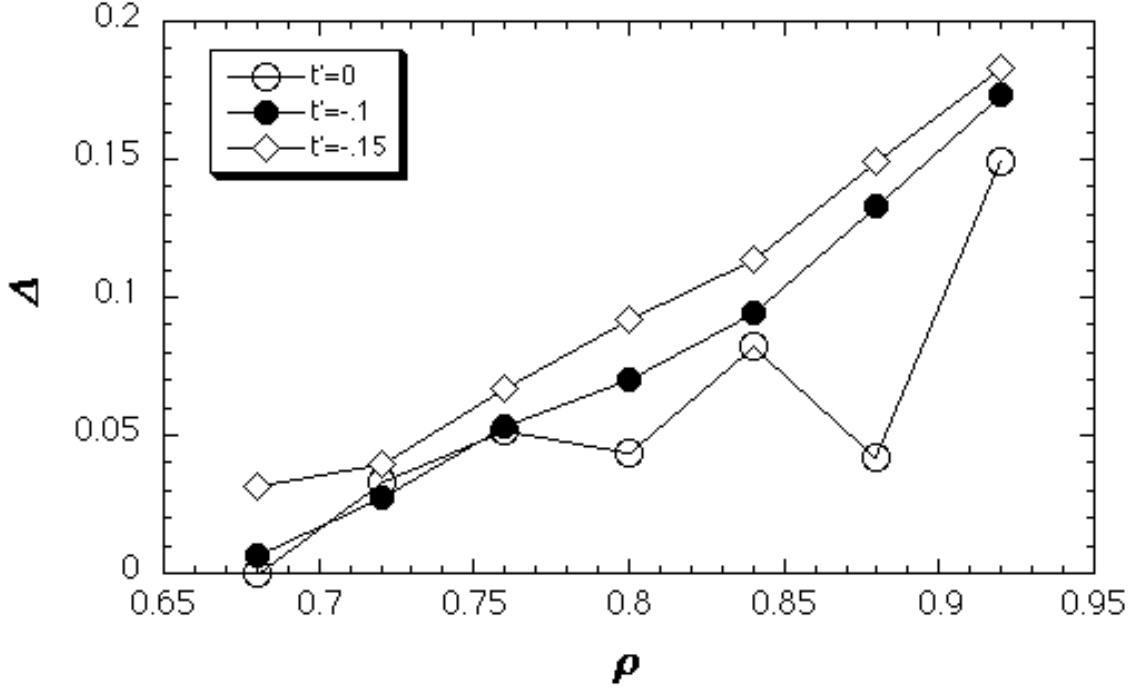

\psbox[width=17cm]{fig10.EPSF}
\caption{Amplitude of the gap function $\Delta$ is plotted 
as functions of $\rho$ for $t' = 0$, $-0.10$ and $-0.15$, 
corresponding to the $\Delta E_g/N_s$ versus $\rho$ curves 
in the preceding figure.
The error bar is about $0.01$ for $\rho \sim 0.92$.
For $\rho \sim 0.7$ it is about 0.005.
}
\label{fig:D-r}
\end{figure}

\subsection{Electron-Density Dependence}

        Figure \ref{fig:gain-r}
shows the energy gain per site in the SC state as a
function of the electron density $\rho = N_e/N_s$ 
for the cases of $t' = 0, - 0.10$ and $-0.15$. 
The corresponding values of $\Delta$ as a function of $\rho$
are exhibited in Fig. \ref{fig:D-r}.
When $t' = - 0.10$ whose curve is exhibited 
with closed circles, the cases of 
$\rho = 0.68, 0.76, 0.84$ and $0.92$ have an open shell, 
in another word, there
are partially filled ${\mib k}$-points
in the electronic state in the $U = 0$
limit, as is shown in Fig. \ref{fig:kpoints} for
$N_s = 10 \times 10$ and $N_e = 84$ with $t' = 0$.
While the other cases 
with $\rho = 0.72, 0.80$ and $0.88$ have 
a closed shell. The curve has almost no dependence
on the shell state in the region of $0.76 \leq \rho \leq 0.92$, 
making a definite contrast with the case with 
$t'=0$ described later. This suggests 
that the curve is already close to
the bulk limit one in this region. 
For $t' = - 0.15$ with open diamonds, the cases of
$\rho = 0.68$, $0.76$ and $0.88$ have an open shell 
while others have a closed shell. 
Here we observe a weak dependence on the shell structure.
However the deviation from the rough average between the 
open-shell and closed-shell curves are only of the order of $1/10$
of the average energy gain per site 
in the region of $0.76 \leq \rho \leq 0.92$, 
which is small enough to judge 
that the points are already close to the bulk limit.
The smooth dependence on $\rho$ is also observed in the SC gap 
amplitude $\Delta$ for $t'=-0.10$ and $-0.15$ in 
Fig. \ref{fig:D-r}.
These results clearly support a statement that the 2D Hubbard model 
with $t' \sim -0.1$ gives a finite condensation energy for the 
$d$-wave SC state in the bulk limit
in the region of $\rho$ from $\sim 0.76$ to $\sim 0.86$. 
For $\rho = 0.72$ a minimum with finite $\Delta$ 
was obtained in both cases of $t' \sim -0.1$. 
But the SC energy gain was within the error bound       
so that $\Delta E_g/N_s \sim 0$.
For $\rho = 0.68$ a finite value was obtained 
for both cases but it is so 
small that it may not remain finite in the bulk limit.

 Open circles for $t' = 0$ in Fig. \ref{fig:gain-r}
form a zigzag curve as a
function of $\rho$. Relatively larger are 
the energy gains for $\rho = 0.76, 0.84$ and $0.92$, 
for which the electron shell is open. 
The values for the cases of $\rho = 0.72, 0.80$ and $0.88$ with
closed shell are smaller. This shell dependence is a system-size effect
and indicates that the $10 \times 10$ lattice 
is not sufficiently large 
when $t' = 0$. It suggests also that the larger the SC energy gain
is, the smaller the sufficient system size is.
However, the average curve between the two curves formed of the
points of both kinds of shell states is expected to be a 
highly probable approximation to the bulk limit and the model
with $t'=0$ is also probable to drive SC, although more extensive
calculation with large lattices is needed to remove the room
for doubt.
Incidentally, for $\rho = 0.68$, no finite energy gain was obtained although the shell is open. 

     In Fig. 1 of the precursive work of
Giamarchi et al. [\citen{rf:gia}] we find a corresponding 
value for the SC energy
gain per site. It is for $\rho = 0.8125$ on the $8 \times 8$
lattice with $U = 10$ and $t'=0$.
Our value for $\rho = 0.80$ on the $10 \times 10$ lattice
with $U = 8$ and $t'=0$ should be close to the value.
The shell structure for $\rho=0.8125$, or $N_e=52$, on the
$8\times 8$ lattice is closed; so is the case 
with $\rho = 0.80$ on the $10 \times 10$ lattice
is closed.
However, our value is $3.38 \times 10^{-4}$, being about 1/3 of the 
above-mentioned value, and if we put it in their figure
it is located in the lower end part of the 
large error bar attached to the data point. 
In the above-mentioned figure the energy gain vanishes around
$\rho \sim 0.56$ but in Fig. \ref{fig:gain-r}
of the present work it vanishes around
$\rho \sim 0.72$. We believe that the present results 
are more improved than the results of Giamarchi et al.
because of improved accuracy of computation.

\section{Competition with the SDW State}

\subsection{Magnetic State of the 2D Hubbard Model}

        When the density of doped holes is small or zero, 
the 2D Hubbard model takes an antiferromagnetic state 
as its ground state. With
increase of doped hole density, 
the magnetic order is destroyed and SC appears. 
At what hole density does the SC state start? 
We have investigated the transition between the pure
SC and the pure uniform SDW states by computing the energy of the SDW state
by means of the variational Monte Carlo method. The trial SDW
wave function is written as [\citen{rf:yoko2,rf:gros}]
\begin{eqnarray}
\Psi_{\mbox{\scriptsize SDW}} &=& P_{\mbox{\scriptsize G}}
                            \psi_{\mbox{\scriptsize SDW}}, \\
\psi_{\mbox{\scriptsize SDW}} &=& 
\prod_{\mib k} (u_{\mib k}c_{{\mib k}\uparrow}^{\dag}
        + v_{\mib k}c_{{\mib k}+{\mib Q}\uparrow}^{\dag})
\prod_{{\mib k}'} (u_{{\mib k}'}c_{{\mib k}'\downarrow}^{\dag}
        - v_{{\mib k}'}c_{{\mib k}'+{\mib Q}\downarrow}^{\dag})
                        |0>, \\
u_{\mib k} &=& [(1 - w_{\mib k}/  
        \sqrt{ w_{\mib k}^2 + M^2 } )/2]^{1/2}, \\
v_{\mib k} &=& [(1 + w_{\mib k}/  
        \sqrt{ w_{\mib k}^2 + M^2 } )/2]^{1/2}, 
\label{eq:vu} \\
w_{\mib k} &=& (\varepsilon_{\mib k} - \varepsilon_{{\mib k} + {\mib Q}} )/2,\\
\varepsilon_{\mib k} &=& \xi_{\mib k} + \mu,
\end{eqnarray}
where $P_{\mbox{\scriptsize G}}$
is the Gutzwiller projection operator defined by eq. (5).
Summation over ${\mib k}$ and ${\mib k}'$ in eq. (20) 
is performed over the filled ${\mib k}$-points,
as in the calculation of the normal state variational M. C. energy.
${\mib Q}$ is the SDW wave vector equal to $(\pi, \pi)$.
$M$ is the SDW potential amplitude.

\begin{figure}
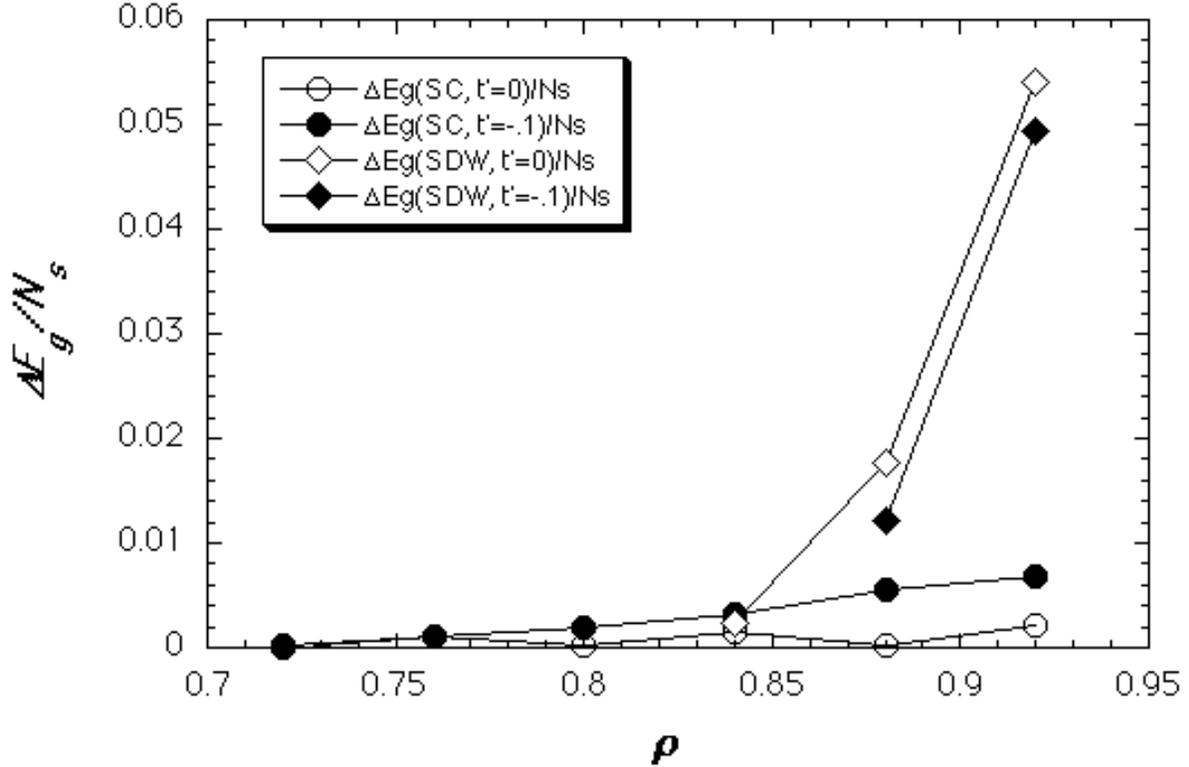

\psbox[width=17cm]{fig11.EPSF}
\caption{Energy gain per site $\Delta E_g/N_s$ in the SDW state against electron density $\rho$ is plotted for $t' = 0$ (open diamond) and for $t' = -0.10$ (closed diamond).  Error bar is about 
$0.0003$. 
$\Delta E_g/N_s$ in the superconducting state is also plotted again with the same symbols as in Fig. \ref{fig:gain-r} with the diminished vertical scale.}
\label{fig:gain-r2}
\end{figure}

\begin{figure}
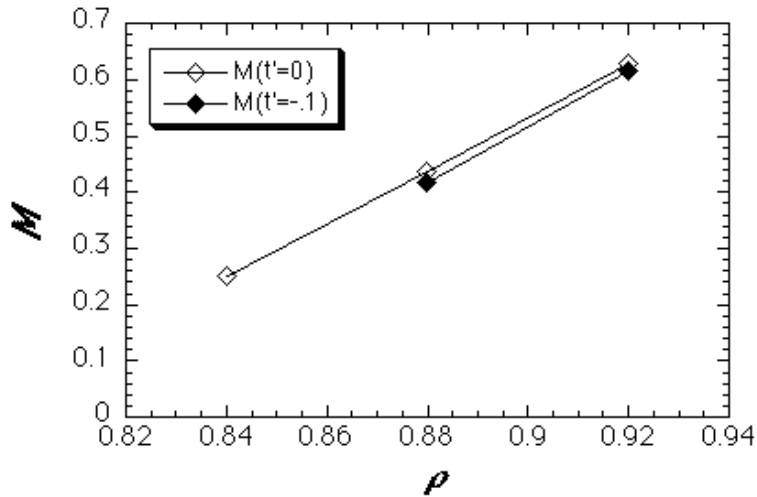

\hspace*{2cm}
\psbox[width=14cm]{fig12.EPSF}
\caption{Optimized value of SDW gap parameter $M$ is plotted as a function of $\rho$ for two values of $t'=0$ (open diamond) and 
$-0.10$ (closed diamond), corresponding to two SDW curves in Fig. \ref{fig:gain-r2}.}
\label{fig:sdwM}
\end{figure}

\subsection{Phase Boundary between SC and SDW States}

        As shown by the open diamonds in Fig. \ref{fig:gain-r2}, 
the energy gain per site in the SDW state rises 
very sharply from $\rho \sim 0.84$ in the case of $t' = 0$. 
Already at $\rho = 0.84$ it takes $0.0023$ 
and is slightly larger than that in the SC state. 
However at $\rho = 0.80$ there is no more stable SDW state. 
By extrapolating the sharply rising curve of the SDW energy gain 
per site as a function of $\rho$, the
boundary between the SDW and the SC states is given at  $\rho = 0.838$.
Optimized values of $M$ are plotted in Fig. \ref{fig:sdwM}. 

\begin{figure}
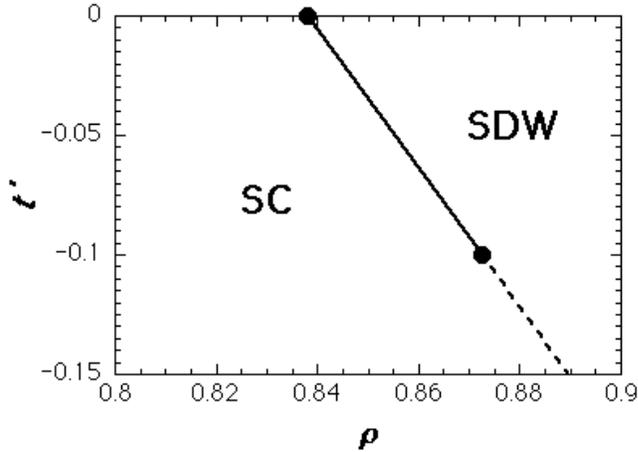

\hspace*{35mm}
\psbox[width=10cm]{fig13.EPSF}
\caption{Phase diagram of superconducting (SC)  and SDW phases in cuprates high-$T_c$ superconductors in the plane of temperature $T$ versus electron and hole densities. The thick curve was decided as written in the text. The dotted line is an extrapolation.}
\label{fig:phasedi}
\end{figure}

        Since finite $t'$ deteriorates 
the nesting property of the Fermi surface, 
the energy gain per site in the SDW state should decrease with
introduction of negative $t'$. The calculated values for 
$t' = -0.10$ is shown
in Fig. \ref{fig:gain-r2} by closed diamonds. 
The energy gain per site in the SC state
given by closed circles increases when $t'=-0.10$. 
The phase boundary between the
two is estimated at about 0.873 as the intersection of the
extrapolation of the line linking the two closed diamonds and the
curve for the SC energy gain per site with $t' = - 0.10$. So the
boundary is pushed up by $\Delta \rho = 0.035$ 
but not enough as to move the boundary
up to $\rho = 0.95$, the observed boundary 
in La$_{2-x}$Sr$_x$CuO$_4$. The obtained phase diagram in the
$t'$-$\rho$ plane is shown in Fig. \ref{fig:phasedi}.

\section{Comparison with Experiments and Other Results}

\subsection{Energy Gain per Site in the SC State}

 By means of the cell-perturbation method,
Feiner et al. were able to reduced the $d$-$p$ model 
into an effective  one-band Hubbard model [\citen{rf:f1}]. 
In their result the transfer energy
$t$ slightly depends on the occupancy on the two related sites. 
Using typical parameter values for the $d$-$p$ model 
$t$ is estimated at 0.51 eV for hole transfer. 
$U$ is estimated at 3.4 eV
so that $U/t = 6.7$. 
With this relatively small $U$ the authors derive 
an appropriate value of exchange interaction constant $J$
for the $t$-$J$ model, appearing later, thanks to the ferromagnetic
interaction term and the Coulomb interaction between
nearest neighbor sites in the effective one-band Hubbard model.
These values of $t$ and $U$ are common for the cuprate high-$T_c$ 
superconductors. 
The $t'$ value was estimated at around $-0.06$ eV, i.e., $-0.12$
in units of $t$, for hole-type superconductors.
They observed that its weak dependence on each cuprate 
is correlated with the values of $T_c$
among the cuprates.
Since the additional spin-dependent ferromagnetic term and nearest
neighbor Coulomb interactions are relatively small, 
the 2D Hubbard model without these additional interaction
but with the above-mentioned
parameter values should give a right magnitude of the SC energy gain if 
it actually gives a finite value.

        For $U = 8$ and $t' = 0$ the 
maximum SC energy gain per site is approximately
given by the value at $\rho=0.84$ in the neighborhood of the SC-SDW 
phase boundary. 
It is
about 0.0010 as the average of the values for both shell states. 
For $U \approx 6.7$ this
value should diminish by a factor 3/4 according to Fig. 
\ref{fig:DEg-U} so that the
calculated maximum SC energy gain per site is 0.00038 eV/site,
if we take $t = 0.51$ eV. 
The value of $t'$ for hole conduction was evaluated by several authors
at, e.g., $-0.06$ eV [\citen{rf:hyb,rf:f1}], $-0.124$ eV (for the La system)
[\citen{rf:toh}], and $-0.17$ eV [\citen{rf:esk}] for hole-doped cuprates.
If we take $t' = - 0.06$ eV $\cong -0.12$ given by Feiner et al. [\citen{rf:f1}], 
the above energy gain should be multiplied by a factor 2.0
and becomes $0.00076$ eV/site. 
Incidentally, this value of $t'$
is in the range of the optimal energy gain, $t'\cong -0.1 \sim -0.15$.

        According to Hao at al. the critical magnetic field 
$H_c(0)$ at zero temperature was estimated to be 1.10 Tesla 
for YBa$_2$Cu$_3$O$_7$ [\citen{rf:hao}]. 
From the expression
$H_c^2/8\pi$ for the condensation energy in the SC state, 
the experimental value for the SC energy gain per site 
is equal to 0.00026 eV per Cu site in layers. 
Triscone et al. [\citen{rf:tris}] gave slightly larger values of $H_c$ for
two samples of YBa$_2$Cu$_3$O$_7$, 
1.231 and 1.364 Tesla,
corresponding to condensation energies $0.00033$ and $0.00040$ eV/(Cu site), respectively.

 Another source of information on the condensation energy is the 
specific heat reported by Loram et al. 
on YBa$_2$Cu$_3$O$_7$ [\citen{rf:loram}].
By numerically integrating the SC specific heat minus the normal-state
one with respect to temperature from zero to just $T_c$, the 
condensation energy was obtained as $0.000168$ eV/(Cu site).
Since the appreciable fluctuation contribution 
is observed at temperatures above 
$T_c$, which is recognized as
the maximum position of the specific heat, 
and since it was simply neglected, 
this value is in fair agreement with the value 
given from the critical field.

 The above-mentioned calculated value 
of the maximum SC energy gain per site
is in reasonable agreement with the 
experimental SC condensation energy in view of simplifications and 
uncertainties in the parameter values of the model. 
This agreement strongly indicates
that the 2D Hubbard model 
includes essential ingredients for the SC in the cuprate 
superconductors.

        The present results of the energy gain per site 
in the SC state
can be compared with the results for the $t$-$J$ model. 
For the value of exchange interaction constant
$J = 4t^2/U = 0.5$ corresponding 
to $U = 8$, Fig. 7 in [\citen{rf:yoko-o}] by
Yokoyama and Ogata allows to estimate the SC energy gain per
site calculated for the $t$-$J$ model. 
At $\rho = 0.84$ it is $0.026t$. 
This is 17 times larger than that obtained 
for the 2D Hubbard model at the same electron density in \S3. 
If we take 0.0010 for the SC energy gain per site 
of the 2D Hubbard model, 
judging from the average of the open and closed-shell
curves, the ratio is 26.
Since the 2D Hubbard model has been found to give a sound 
SC condensation energy as seen above, we have to judge that the 
fault is on the 2D $t$-$J$ model as an effective Hamiltonian 
of the 2D Hubbard model, i.e., it gives too large an SC
condensation energy at least in the parameter region where
$J/t \sim 1/2$. This means that the $t$-$J$ model made of the 
leading two terms in the expansion in powers of $t/U$ of the
canonical transformation of the Hubbard model should be treated 
together with the higher-order terms for it to give a realistic 
SC condensation energy. 
 
 The $t$-$J$ model
is derived also starting from the $d$-$p$ model or three-band Hubbard
model. As such its parameter values are derived in literatures.
The values are given in wide ranges, even by a single group.
The value of $t$ is estimated at $0.224$ eV by Tohyama et al.
for the La system [\citen{rf:toh}],
at $0.51$ eV by Feiner et al. [\citen{rf:f1}], in the range of $0.29 \sim 0.98$ eV by
Batista et al. [\citen{rf:bat}]. The value of $J$ is obtained at $0.13$ eV for 
La$_2$CuO$_4$ from Raman scattering data [\citen{rf:toku}]. Theoretical values given
by these authors for typical cases are around this value. If we take
$J=0.13$ eV and $t=2J$, the SC energy gain of the 2D $t$-$J$ model with
$\rho=0.84$ is estimated at $0.026t=0.0068$eV/site by using [\citen{rf:yoko-o}];
it is 26 times larger than the value obtained from Hao et al.'s $H_c$ 
value. Around this value of $J/t=0.5$ the SC energy gain of the 
$t$-$J$ model is considered to be roughly proportional to 
$t \cdot J/t =J$, if the SC energy gain 
of the model with normalized coupling constant $J/t$
is roughly proportional to $J/t$. 
Then, the above-mentioned large ratio applies to the
most parameter sets. Only when $J/t$ becomes so small, 
e.g., with increase of $t$,  that SC almost vanishes, 
the above ratio may decrease to unity. Thus with plausible
parameter sets, the $t$-$J$ model is very probable to give too
large a SC condensation energy. Again the higher order correction 
terms must not be neglected, for the model to work properly at least
concerning the SC condensation energy, 
so that they oppose to the occurrence to SC [\citen{rf:bat}]; 
the present result does not support 
the correction terms of the type that facilitate SC [\citen{rf:matsu}].

\subsection{Phase Diagram in the $T_c$-versus-$\rho$ Plane}

        According to our results in \S4, in the case of small negative $t'$ the SC region extends
from $\rho \sim 0.76$ to 
$\rho \sim 0.86$ with smoothly increasing SC condensation energy
with increase of $\rho$.
This suggests that in this $\rho$ region $T_c$ is finite and 
smoothly increases with increase of $\rho$.
The range of this $\rho$ region and the expected smooth increase 
of $T_c$ are in fair agreement with 
the features of the $T_c$-versus-$p$ 
($p = 1 - \rho$, doped-hole density) 
phase diagram in the optimal to overdoped region, 
i.e., $0.15 \leq p \leq 0.25$ or $0.75 \leq \rho \leq 0.85$, 
where $T_c$ increases with increasing $\rho$
[\citen{rf:taka}].

Corresponding to the underdoped region with
$ 0.05 \leq p \leq 0.15$, 
the present results do not give an SC region
but give an SDW region. However, there is a possibility
that more elaborate calculations give a kind of SC phase where 
SC and SDW coexist and that the phase diagram
in the underdoped region is also provided by the 2D Hubbard model.
The reasons are as follows:  
Giamarchi et al. showed that a uniform wave function describing
the coexistence of SC and SDW gives a lower energy than the pure
SDW wave function in the system of 32 electrons on the $6\times6$
lattice [\citen{rf:gia}]. We have checked on the $10\times 10$ lattice
that such a coexistence wave
function is possible only in the $\rho$
region higher than the above-mentioned boundary between SC and SDW.
Therefore, in the higher $\rho$ region 
there may be a coexistence phase in a certain region. 
However, there are no experiments confirming a uniform
coexistence except for recent reports suggesting a non-uniform
coexistence in La$_{1.6-x}$Nd$_{0.4}$Sr$_x$CuO$_4$ [\citen{rf:tra}].
Further, stripe-type solutions are known to give
a lower total energy for the SDW state in the low hole-doping region [\citen{rf:gia2}].
Therefore, the above-mentioned uniform coexistence state
may be dominated by a more complicated non-uniform way of 
coexistence of SC and SDW.
Such a ground state could be obtained only if we use sufficiently
large sizes of the system, as is the case with the 
SDW stripe solutions [\citen{rf:gia2}].
        With this speculated coexistent SC state, 
we expect that $T_c$ is maximum at the
above-mentioned phase boundary at $\rho \sim 0.86$,
since 
the plausible coexistence state should have $T_c$ lower
than the value of $T_c$ at the boundary
due to a smaller electron density carrying
SC properties. 
At temperatures higher than $T_c$, the SDW ordering 
without the SC one may remain.
Such a coexistent SC state may be 
consistent with the feature 
of the $T_c$-$p$ phase diagram in the underdoped region [\citen{rf:taka}]. Of course, all these possibilities remain to be confirmed with actual calculations.

\subsection{Comments on the Quantum Monte Carlo Results}

 Whether the 2D Hubbard model drives SC or not has been an important
unsettled issue for theoretical and computational physics. Although 
quantum M. C. studies recognized the enhancement of the SC
susceptibility, dominance of SC has not been confirmed.
However, the numerical studies have been under severe restrictions
to the parameter values in which reliable results were obtainable, e.g., with respect to the system size, value of $U/t$, temperature and so on. Within these few years the occurrence of the SC was strongly indicated 
when $t'$ is introduced [\citen{rf:huss,rf:asai}]. 
When the shell structure approaches to the open 
one, even if it was closed, SC correlations were observed to be much 
enhanced [\citen{rf:kuro}].  On the contrary, 
recently Zhang et al. showed that SC features weaken with increase of the system size with $U$ up to $8t$ [\citen{rf:zha}].
However, the state with $\rho = 0.85$ and $t' = 0$
with which their main results are exhibited
is located in the SDW region according to our results. 
With this value of $\rho$
the dominant SDW is considered to have diminished the SC
correlations for some values of $U$.  
Even in the case where the coexistence of SC and SDW 
might have occurred,
the SC features should have been different from what we expect in the usual 
uniform superconducting state. 
Therefore, their results do not disprove the existence of SC in the 
$\rho$ region where we find SC dominating. 
Further the quantum M.C. calculations
were done for closed-shell electron densities.
This shell state must have led to a very small energy gain 
and weakened the SC features when the system size is not sufficiently large.
The distribution of the single particle energy $\varepsilon_{\mib
k}$ around the Fermi energy are scarce
in the cases they treated.
These reasons are suspected to have made the SC features in
their work very weak. A study with an elaborate method allowing high
precision is under way searching enhanced SC correlations with appropriate
values of $\rho$ [\citen{rf:yana}].

\section{Summary}

        Using the variational Monte Carlo method we have investigated
the dependencies of the energy gain per site in the SC state on
important parameters such as on-site Coulomb energy $U$, 
next nearest neighbor transfer energy $t'$ and 
electron density per site $\rho$.
We worked mainly with the $10 \times 10$ lattice. 
The energy gain is maximized at about $U = 8$. 
It is maximized for $t' = - 0.10 \sim -0.15$,
reaching twice the value at $t' = 0$. 
This increase of the SC energy gain was ascribed
to the fact that a high density of the one-electron levels around the van Hove
singularity go close  to
$\varepsilon_{\mbox{\scriptsize F}}$ due to an appropriate
negative value of $t'$.
The SC pair correlation functions increased correspondingly. 
With $U = 8$ and $t' = - 0.10$ or $-0.15$, 
the energy gain starts to be finite at 
$\rho \sim 0.68$,
smoothly increasing with increase of $\rho$ 
from $\rho = 0.76$
up to $\rho = 0.92$ up to
which we have made calculation. 
The curves of the energy gain as a function of $\rho$
have only a weak
dependence on the shell structure or on whether some of 
the ${\mib k}$-points are partially
filled or not (open or closed shell) in the $U = 0$ limit electron
distribution. This definitely suggests that the curves are already 
close to the bulk limit ones. 
On the other hand, with $U = 8$ and $t' = 0$, the energy gain 
in the open shell state is sizable 
but that in the closed shell state is rather suppressed
so that the obtained values seem to have still appreciable 
system size dependence,
although the average curve between the open- and 
closed-shell curves is
considered to be a fair approximation to the bulk limit. 
We have also
calculated the energy gain in the commensurate SDW state. 
It sharply rises as a function of $\rho$ starting at 
about $\rho \sim 0.84$ for $\rho = 0$, making the SDW state the lower
energy state in the higher $\rho$ region above 
this value. 
A finite value of $t' \sim -0.1$ brought about an upward shift of the boundary by $\Delta t' \sim 0.035$. 
Therefore, our calculations confirmed that the electronic
repulsive interaction can drive SC by itself in the region from $\rho \sim 0.76$ up to the boundary at $\rho \sim 0.87$ with $t' \sim -0.1$.
The maximum SC energy gain per site in the SC region was 
found to be close to the experimental SC condensation energy
evaluated from $H_c$ and the specific heat of YBa$_2$Cu$_3$O$_7$. 
On the other hand, the corresponding $t$-$J$ model 
was pointed out to give a value
larger enormously by a factor exceeding 20, 
indicating that it is not quantitatively reliable as 
a model for high-$T_c$ cuprates
without taking account of higher-order correction terms. 
The feature of the experimental phase diagram 
in the $T_c$-$\rho$ plane 
in the region of $\rho$ lower than the $\rho$ of the
maximum $T_c$ was asserted to be in a  good
agreement with our results. 
These results strongly suggest that the simple 2D Hubbard model
includes essential ingredients of high-$T_c$ cuprates.
A speculative argument on the higher $\rho$ region outside 
our SC region was given concerning the coexistence of SC and SDW.

\section*{Acknowledgments}

The authors are cordially thankful to Dr. T. Giamarchi, 
Professor J. Kondo, Dr. K. Kuroki, Professor T. Moriya,
Professor M. Ogata and Dr. R. Ramakumar for valuable discussions.

\end{document}